\documentclass[preprint,12pt]{elsarticle}
\usepackage{amssymb}
\journal{Physics Letters B}
\usepackage{epsfig}
\usepackage{graphicx}
\usepackage{color}
\usepackage{soul}
\usepackage{ulem}
\usepackage{amssymb}
\def\ee{\end{equation}}
\def\be{\begin{equation}}
\def\eea{\end{eqnarray}}
\def\bea{\begin{eqnarray}}
\def\pp{$pp$}
\def\pbarp{{\bar p}p}

\def\sigtot{\sigma_{\rm total}}

\def\dsigdt{d\sigma_{\rm el}/dt}

\def\Reo{\Re e}
\def\Imo{\Im m}
\def\tev {\ensuremath{\rm TeV}}
\def\gev {\ensuremath{\rm GeV}}
\def\mb {\ensuremath{\rm mb}}

\def\sigeltosigtot{$\sigma_{\rm elastic}/\sigma_{\rm total}$}
\begin{document}

\begin{frontmatter}
%\documentclass[12pt]{article}

%%%%%%%%%%%%%%%%%%%%%%%%%%%%%%%%%%%%%%%%%%%%%%%%%%%%%%%%%%%%%%%%%%%%%%%%%%%%
%  personal abbreviations and macros
%    the following package contains macros used in this document:
%\input econfmacros.tex
%%%%%%%%%%%%%%%%%%%%%%%%%%%%%%%%%%%%%%%%%%%%%%%%%%%%%%%%%%%%%%%%%%%%%%%%%%%
%
%  To include an item in the INDEX of the conference volume,
%           flag it with    \index{<item name>}
%  The use of this macro is illustrated in the text.
%
%%%%%%\includegraphics[]{dipxplbjune4final.pdf}
%%%%%%%%%%%%%%%%%%%%%%%%%%%%%%%%%%%%%%%%%%%%%%%%%%%%%%%%%%%%%%%%%%%%%%%

%%%%%%%%\def\Title#1{\begin{center} {\Large {\bf #1} } \end{center}}
%%%%%%%%%%\usepackage{fancyhdr}
%%%%%%%%%\textheight = 609pt 
 %%%%%%%%%\textwidth = 484pt
%%%%%%%%%%%\setlength{\oddsidemargin}{-0.3cm}
%%%%%%%%%%%\pagestyle{fancy}
%%%%%%%%%%\cfoot{\thepage}
%\rhead{Frascati Physics Series - Vol. LIII}
%\lhead{LC11 Proceeedings}
%%%%%%%%\begin{document}
%%%%%%%%%\topskip 2cm 
%%\setcounter{tocdepth}{5}
%%%\tableofcontents
%%%\setcounter{secnumdepth}{5}

%%%%%%%\Title{
%Checking asymptotia at LHC through pp elastic scattering in  a model independent description}
%pp cross-sections: a QCD model compared with TOTEM and other
%LHC  data}
%An alternative title: 
\title{Checks of asymptotia in $pp$ elastic scattering at LHC}
%
%%%%%%%\bigskip

%%%%%%%%\begin{raggedright}

%\vspace{0.2cm}
\author[label1]{ Agnes Grau} 
\address[label1]{Departamento de F\'\i sica Te\'orica y del Cosmos,Universidad de Granada, 18071 Granada, Spain}
%\address[label2] {igrau@ugr.es}
%\author[label2]{Rohni M. Godbole}
%\address[label2]{Centre for High Energy Physics, Indian Institute of Science, Bangalore, 560 012, India}
\author[label5]{ Simone Pacetti}
\address[label5]{INFN \& Physics Department, University of Perugia, Italy}
%\address[label6]{simone.pacetti@pg.infn.it}
\author[label6]{Giulia Pancheri}
\address[label6]{INFN Frascati National Laboratories, Via E. Fermi 40, Italy}
%\address[label8]{pancheri@lnf.infn.it}
\author[label5]{Yogendra N. Srivastava}
%\address[label9]{yogendra.srivastava@pg.infn.it}

\begin{abstract}
{We parametrize  TOTEM data for the elastic differential \pp \ cross section at $\sqrt{s}=7\ \tev$ 
in terms of two exponentials with a relative  phase. 
We employ two previously derived sum rules 
for $pp$   elastic scattering amplitude in impact parameter space 
to check whether asymptotia has been reached at the LHC. 
A detailed study of the  TOTEM data for the elastic differential cross section at $\sqrt{s}=7\ \tev$ is made 
and it is shown that, within errors,  the asymptotic sum rules are satisfied at LHC. 
We propose to use this parametrization to study forthcoming higher energy data.}
\end{abstract}
\begin{keyword}
proton-proton \sep elastic scattering \sep total cross section 
\end{keyword}

\end{frontmatter}
%\vskip 1cm
%{\it The paper as is will still have to be  be reduced by 1/4, I have taken away the things in italics and sout, the longer versions are pre-may5th, what follows is the version to insert in the Phys. Lett. format}
%\vskip 1 cm]
\section{Introduction}
The TOTEM experiment has measured the total {\pp}\ cross section at $\sqrt{s}~=~7~\tev$~\cite{totem-total} and the value   may be consistent  \cite{bhasymptotia} with saturation of the { Froissart-Martin} bound  \cite{froissart,martin}, {\it i.e.} 
$\sigma_{\rm total}\simeq \ln^2 s$. It is also consistent with a range 
of total cross section predictions, based on a mini-jet QCD model with infrared gluons~\cite{ourplb08,ourPRD11}, as well other 
%predictions 
deductions based on a hard Pomeron component~\cite{DL2012}. The saturation of the Froissart bound,  if confirmed, could  exclude, at least at present energies~\cite{yogi-hidden-extra},  a different behaviour  obtained from hidden extra dimensions~\cite{Amati,Agashe}. Indeed, according to {Ref.}~\cite{bhhidden}, the TOTEM result indicates that we have now reached asymptotia  as far as the total cross section is considered  and thus  existence of  hidden extra dimensions can be excluded. On the other hand, contradictions in   the analysis of Ultra High Energy Cosmic Rays composition~\cite{tsvi} might hint to the existence of new physics that modifies  baryonic interactions  at {center-of-mass (CM)} energies  in the $50\div 100 \ \tev$ range. Thus the problem of the energy dependence of the total cross section is 
%still open. 
not yet beyond all dispute.

 Another { interesting feature}  has arisen in connection with TOTEM  published results for the differential  and the total  elastic cross section~\cite{totem-elastic}. These data  show that  the dip, which had characterized data at ISR in $pp$ scattering, and which has been only a faint presence in $\pbarp$ 
at the TeVatron, has reappeared.  We propose to use these data to check whether asymptotia has been reached, through two asymptotic sum rules (ASR) for the 
elastic amplitude in impact parameter space~\cite{Pancheri:2004xc}.
%\cite{Pancheri:2005jr}. 
To check the satisfaction of these ASR, we shall use { the}{ model independent 
parametrization of the scattering amplitude~\cite{bp1973}  proposed in 1973 by Barger and Phillips.
 %\cite{bp1973}. 
 
 { We shall see that t}his model, with two exponentials and a phase, 
 %\sout{ is the simplest possible phenomenological model through which we can investigate the structure of the dip.  We shall see that this  mode}
 describes  well both the   diffraction peak, 
 as well as the $t$-dependence after the dip at LHC {with} CM energy of 7~\tev\ (LHC7). 
We also analyze ISR data for $pp$ scattering at  $\sqrt{s}=53\ \gev$, and, by comparison,  find that the ASR
are  not satisfied   at ISR,  while  being almost exactly satisfied at LHC7.  We  discuss   the  approach to asymptotia of the ratio 
 \sigeltosigtot\ and of  the forward slope and  suggest to use this simple model to fit  future forward scattering LHC data.

\section{ Asymptotic sum rules for the elastic scattering amplitude}
\label{sec2}
Let   the elastic amplitude $F(s,t)$ be normalized  so that 
\bea
\sigma_{\rm total}(s) = 4 \pi\, {\Imo}\,  F(s,0) \,.
\label{2}
\eea
At high energy, all particle masses can be ignored and the elastic differential cross section reads  
\bea
\label{2}
\frac{d\sigma}{dt} = \pi |F(s,t)|^2\,.
\eea
%\sout{In terms of the impact parameter representation,}
 { Writing the elastic amplitude}
 % \sout{ can be written} 
 as 
\bea
\label{3}
F(s,t) = i \int_0^\infty (bdb) J_0\left(b\sqrt{-t}\right) 
\left[1 - e^{2i\delta_R(b,s)}e^{-2\delta_I(b,s)}\right]\,,
\eea
%\ctr{where: $J_0$ is the $0^{\rm th}$ order Bessel function and, 
%$\delta_R(b,s)$ and $\delta_I(b,s)$ are the real and imaginary
%parts of the complex phase shift $\delta(b,s)$.}
%%where $\delta(b,s)$ denotes the complex phase shift. 
inversion of Eq.~(\ref{3}) reads
\bea
\label{4}
\left[1 - e^{2i\delta_R(b,s)}e^{-2\delta_I(b,s)}\right] = 
-i\, \frac{1}{2} \int_{-\infty}^0 (dt) J_0\left(b\sqrt{-t}\right) F(s,t)\,,
\eea
{ and }
 %\sout{Using the eikonal representation and its correspondence with  the partial wave expansion, } 
 one can study properties of the amplitude in impact parameter space from the known behaviour in the angular momentum discrete variable. The Froissart bound implies that there must exist  a finite angular momentum value, {\it below} which all partial waves are  absorbed. { Under the stronger hypothesis of}
%Using the semiclasical relation between the angular momentum and impact parameter $l=bk$ 
%\sout{This leads to}
 total absorption as $b\rightarrow 0$ in the ultra high energy limit, namely
%We know from our work and as well as from that of the others, that all
 all ``low-$b$'' waves are { completely} absorbed, { we have}
 % viz., that ${\Im m}\ \eta(0,s)\rightarrow\ \infty$ as $s\rightarrow\ \infty$. 
 %\sout{This gives us} 
 the two ASR
\bea
\label{5}
SR_1=\frac{1}{2} \int_{-\infty}^0 (dt) {\Imo} F(s,t)& \rightarrow&  1;\ \mbox{as }\ s\rightarrow\ \infty\,,\\
SR_0=\frac{1}{2}\int_{-\infty}^0 (dt) {\Reo} F(s,t) & \rightarrow &  0;\ \mbox{as }\ s\rightarrow\ \infty\,.
\eea
Satisfaction of these rules is a good measure of whether the asymptotic limit has been reached, and would reinforce the statement~\cite{bhasymptotia}  based on  the TOTEM data for  total cross section, that we 
may have reached asymptotia, namely that  $\sigma_{\rm total}\sim \ln^2 s$,   and 
 saturation of the Froissart bound has been observed.  
  { Notice that the Froissart-Martin bound is obtained under a  hypothesis weaker than total absorption \cite{martin} and it would lead to $SR_1\to\ 2$; $SR_0\to\ 0$. As we shall see, phenomenologically Eq.(\ref{5}), is favoured.}
  
% Checling of
 In order to check these ASR, one requires a model for the scattering amplitude.
%\subsection{Real and imaginary parts of the elastic scattering amplitude-april 28}
While the imaginary part of the elastic amplitude is solidly anchored to the optical theorem, the real part is more model dependent. A limiting behaviour for the real part of the forward amplitude has been obtained by  Khuri and Kinoshita~\cite{khuri}, {\it i.e.}, for $s \rightarrow \infty$
\bea
\rho(s,0) = \frac{\Reo F(s,0)}{\Imo  F(s,0)} \approx  \frac{ \pi} {\ln(s/s_0)}\,,
\eea
%with $constant =\pi$ 
if the Froissart bound is %satisfied. 
saturated. As discussed later, the model of Ref.~\cite{ourfroissart} 
gives  $\rho(s,0)=\pi/[2p \ln(s/s_0)]$ with $p=1$ for a total cross section asymptotically 
growing as $\sim \ln s$, and $p=1/2$, when the Froissart bound is saturated. 

For values of $q^2=-t\neq 0$, there is a 
%heuristic, bur rather general  
%rather general albeit heuristic
result by Andr\'e Martin~\cite{Martin:1973qm,Auberson:1971ru}, which relates the real part of the amplitude to its imaginary part. Given an $\Imo F(s,t)$, Martin's  method 
%allows one to construct 
gives an asymptotic expression for $\Reo F(s,t)$, i.e. 
\bea
\Reo F(s,t)=\rho(s,0)\frac{d}{dt}\big[t\, \Imo  F(s,t)\big]\,.
\eea
We note here
%without proof, 
that 
%a simple derivation such as the one above
the above equation is valid also if  the Froissart bound is not saturated, but 
$\sigma_{\rm total}\sim \ln^{1/p}(s)$, where  $1/2\le p\le1$. 
%Then, 
The elastic amplitude still scales with the variable 
$\tau=t\, \sigma_{\rm total}$,  the only difference is
%as shown below,  
$\rho=\pi/\big[2 p \ln (s/s_0)\big]$, {with $s_0=1\,\gev^2$}.

To complete the expression for the real part, one needs to estimate $\rho(s,0)$ for all cases including the case when the Froissart bound is not saturated.
% To obtain 
%in \ctr{Ref.}~\cite{khuri},s 
%the asymptotic expression, it is not necessary to assume the  scaling in the variable $\tau$. 
The phase of the leading (first term) contribution at $t=0$  is readily obtained from the $s$-dependence of the total cross section, or equivalently  $\Im m F(s,0)$, using the prescription $s\rightarrow se^{-i\pi/2}$, 
%For small $t$-values, this can be done by following the Block and Cahn prescription  to make the substitution $s\rightarrow s e^{-i\pi/2}$, which can  
which assures analyticity and crossing symmetry of the scattering amplitude~\cite{Martin:1973qm,blockcahnrep}. 
%at $t=0 $.  Since $a_Y(s)=\sigma_{total} /4 \pi$, Let the forward elastic amplitude be $A(s,0)$. Then, at 
At high energies {\it and very small $t$ values},
%$s-u$  
$s\!\leftrightarrow\!\! u$ crossing symmetry tells us that the leading $C=+$ amplitude is 
a function of the complex variable $s e^{- i \pi /2}$. The Froissart bound requires $\Im m 
F (s,0)\lesssim \ln^2s$, and we  can generally write, asymptotically
\bea
\label{r1}
F (s,0)\rightarrow\ i\left[ \ln\left(\frac{s}{s_0}e^{ - i \frac{\pi}{2}}\right)\right]^{1/p}=
%[ \ln \left(\frac{s}{s_0} e^{- i \pi /2}]^{1/p}\right)= 
i\left[ \ln\left(\frac{s}{s_0}\right) - i\, \frac{\pi}{2}\right]^{1/p}\,,
\label{eq:phase block}
\eea
with  $1/2 \le p\le1$ to include both the case of saturation of the Froissart bound and a slower rise, compatible with $\ln s$ behaviour. 
For large $s$, this may be approximated to
\bea
\label{r2}
F (s,0)\rightarrow\ i\,\left[ \ln\left(\frac{s}{s_0}\right)\right]^{1/p} \left[1 -\frac{ i \pi }{2p \ln(s/s_0)}\right]\,.
\label{eq:phase}
\eea
The above equation gives an estimate for  the 
leading contribution to the parameter $\rho(s,0)$,
%=\Re e {\cal A}(s,0) /\Im m {\cal A}(s,0)$ , 
namely
\bea
%%\label{r3}
\rho(s,0) = \frac{\Reo  F (s,0)}{\Im m F(s,0)} \approx  \frac{\pi}{2p \ln(s/s_0)}\,,
\label{eq:rho}
\eea
where the approximate sign refers to the fact that the real part of the 
amplitude can receive also (non-leading) contributions  from other terms. 
The Khuri-Kinoshita result, valid when the Froissart bound is saturated, 
is obtained for $p\ =\ 1/2$.
%~\cite{khuri}.
 To obtain a numerical value for $p$, 
%SR requests that there are two contributions to the real part of the amplitude, one coming from the first term and one from the second term. For what concerns the first term, which dominates at small $t$,  the addition of a real part may be attempted via the following simple model as follows:
%\\
%\bea
%\label{7}
%\frac{d\sigma}{dt} = [\frac{d\sigma}{dt}]_{Im, t=0} [1+ \rho(s,t=0)^2] e^{Bt}  \ \ \ \ \ \ t \sim 0
%\eea
%with
%\bea
%\label{10}
%\rho\ =\ \frac{\Re e A(s,0)}{\Im m A(s,0)}\ = \frac{\pi}{2p ln(s/s_o)},
%\label{eq:rho}
%\eea
%and $ [\frac{d\sigma}{dt}]_{Im, t=0} $ is the differential elastic cross section at $t=0$ computed only using the imaginary part of the amplitude. 
%To understand Eq. (\ref{eq:rho}) one can use the argument by Block and a demonstration by Yogi.
%\section{The $\rho$ parameter \label{rho}}
%A value for the parameter $p$ can be obtained from a model for the total cross section 
% well as the because $\pi /2p ln(s/s_o) $ is not so small at lower energies and the approximation  in Eqs.~(\ref{eq:phase}) and (\ref{eq:phaseapprox}) are 
%non valid. .
we use a model~\cite{ourPRD11} built on QCD mini-jets embedded into the eikonal representation. According to this model, QCD mini-jets   drive the rise of the cross section, and  $k_t$-resummation of ultra-soft gluons  transforms the power-law rise of the mini-jet cross sections into an asymptotic logarithmic behaviour.  In our model 
%we find
~\cite{ourfroissart} 
\bea
\sigma_{\rm total}\sim \ln^{1/p}(s) \,,
\eea
where the parameter $1/2<p<1$ can be related to a dressed  one-gluon exchange potential rising 
%like 
as $r^{2p-1}$,  and our phenomenology of total cross section data~\cite{ourplb08} leads to 
values   $ p=0.66\div 0.77 $. 
%The above expression can be  used  when constructing the scattering amplitude in impact parameter representation. In such cases, the derivative introduces an extra term which fills the zero of the imaginary part of the scattering amplitude.

%\section{Elastic scattering  and asymptotic sum rules}

 It should be noted that the asymptotic  expression for the real part of the scattering amplitude automatically satisfies the second ASR. However, this does not help in checking whether asymptotia has been reached, since the expression for the real part is based on an asymptotic scaling law, which may not be valid.
 
%\section{The rise with energy of $\sigma_{elastic}$}
{ Another  measure for asymptotia, also examined in {Ref.}~\cite{bhasymptotia}, can be obtained from the ratio of the elastic to the total cross section.  Based on the black and grey disk model, where $\sigma_{\rm total}=2\pi A R^2(s)$ and $\sigma_{\rm elastic }=\pi A^2 R^2(s)$, the black disk limit, $A=1$, gives, for the ratio,  the limiting value  1/2. However, we have recently shown~\cite{ourPRD11} that the eikonal formulation at high energy underestimates  the total inelastic cross section, including in it only uncorrelated (non-diffractive) collisions, whereas the diffractive collisions appear to be part of the eikonal elastic cross section. In fact, it is  known that, at high energies, 
%As a conseguence, 
the ``elastic" cross section, as defined in eikonal models, is typically  overestimated~\cite{lipari}.    Thus we propose the following asymptotic limit
\bea
{\cal R}= \frac{\sigma_{\rm elastic}+\sigma_{\rm diffractive}}{\sigma_{\rm total}}
= {\cal R}_{\rm el}+{\cal R}_{\rm diff}\rightarrow 1/2\,, \ \ \ \ s\rightarrow \infty\,,
\label{eq:ratio}
\eea
where $\sigma_{\rm diffractive}$ includes both single and double diffractive components. 
We show in Fig.~\ref{fig:sigeltosigtotnew} a compilation of data ~\cite{pdg} for the ratio 
$\sigma_{\rm elastic}/\sigma_{\rm total}$ and compare it with the black disk 
limit and results of {Ref.}~\cite{bhasymptotia}. 
%While the lower energy data are compatible with a logarithmic rise, data from the TeVatron to the LHC give a conflicting behaviour, which is influenced by the more than 10\% uncertainty at the TeVatron.
\begin{figure}[h!]
\centering
%\hspace{-0.5cm}
\resizebox{0.8\textwidth}{!}{
\includegraphics{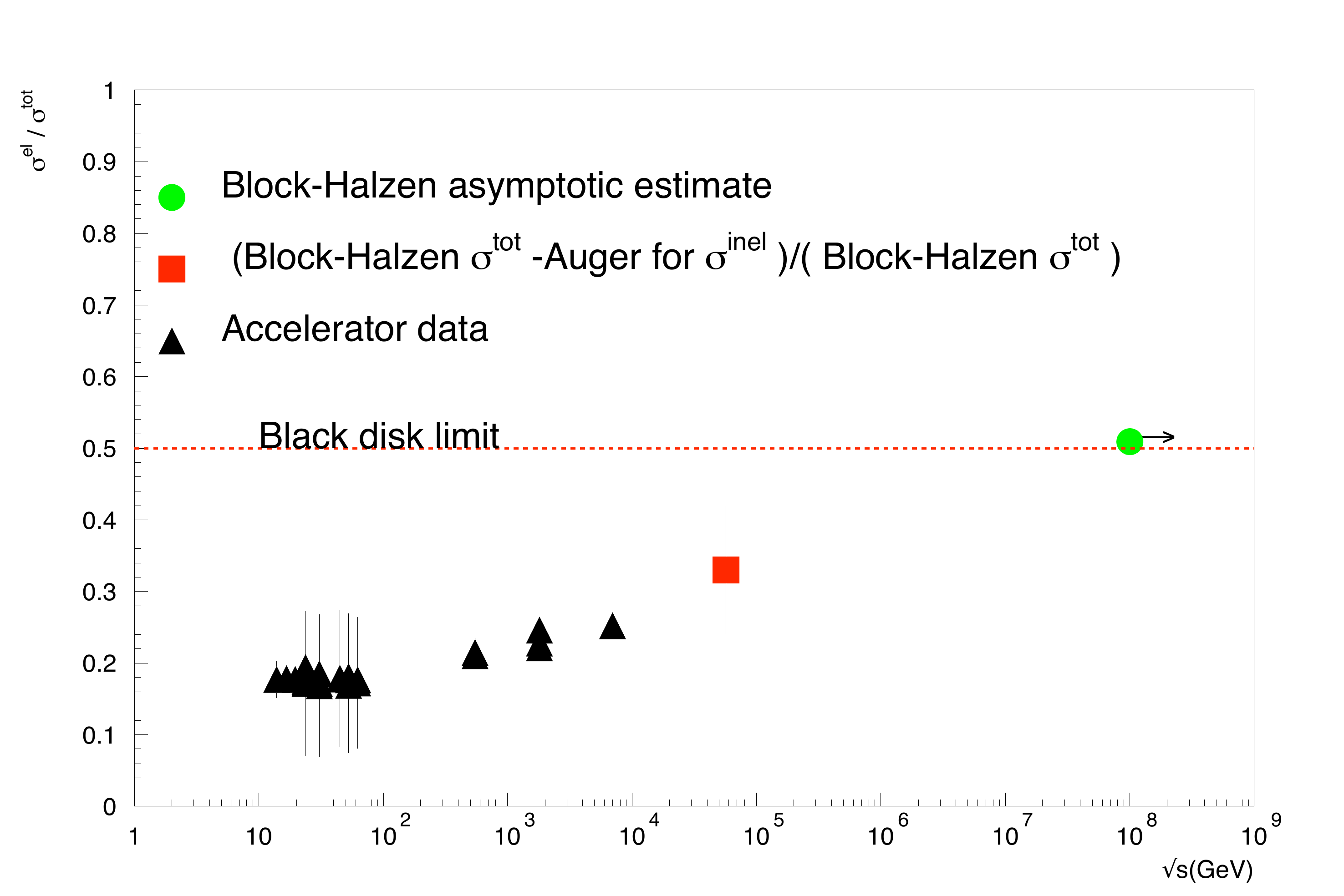}}
%\resizebox{0.4\textwidth}{!}{
%\includegraphics{simone-el-tot}}
\caption{\label{fig:sigeltosigtotnew} The ratio \sigeltosigtot\ 
compared with accelerator data (black triangles)~\cite{pdg}, including TOTEM.
% up to $\sqrt{s}=7\ \tev$. 
The value at $\sqrt{s}=57\ \tev$ is obtained from the ratio of the Auger
collaboration value for $\sigma_{\rm inel}$ { \cite{auger2011}} to the value for $\sigma_{\rm total}$ 
predicted in {Ref.}~\cite{bhasymptotia}. }
%An enlargement of the accelerator data is shown at right.{\it This two figures will have to be redone, at left I only used a few values for ISR, also the green dot for BH is really an asymptotic limit not a point. The two  figures  will have to be done with similar style and the one at right needs labelling of the points.}}
\end{figure}
% It is just as possible, that the logarithmic behaviour of the total elastic cross section be different from that of the total cross section [???]. 
 %In a model where the cut-off in impact parameer spece is determined by soft gluon behaviour in the infrared, 
In the figure we have added to the accelerator data
% for the ratio \sigeltosigtot  
a   value  extracted from the recent  measurement by the Auger Collaboration for the inelastic cross section ~\cite{auger2011}. 
We estimated the ratio ${\cal R}_{\rm el}$ at 57 \tev, by using Block and Halzen (BH)~\cite{bhasymptotia} 
value for the total cross section at $\sigma^{\rm BH}_{\rm total}(57\ \tev)=(134.8\pm 1.5)$ mb,  which is based on the analytic amplitude method  of {Ref.}~\cite{bhamplitude}. We then obtained 
$\sigma_{\rm elastic}(57\ \tev)=\sigma^{BH}_{\rm total}-\sigma^{\rm Auger}_{\rm inelastic}=(44.8\pm11.6) \ mb$. We also show the asymptotic result  (green dot)
from {Ref.}~\cite{bhasymptotia}, which  appears in line with the ratio \sigeltosigtot \ 
%${\cal R}$ 
extrapolated from the  Auger point for the inelastic cross section. The conclusion in {Ref.}~\cite{bhasymptotia} that we are still not in asymptotia for what concerns this ratio, may however be too pessimistic.  Firstly, from the figure, it is quite conceivable that data may approach a limiting value of 1/3.  This would be better in keeping with Eq.~(\ref{eq:ratio}). Also, 
%In addition, there are two related considerations which can affect the value we obtain  at 57 \tev.   
estimates for the inelastic cross section are  affected by large uncertainties, and often model dependent on extrapolations. Indeed, in order  to reconcile cosmic ray composition  data,   an   extrapolation from  
$p$-air to \pp\  could use    higher total  cross section values~\cite{tsvi}. Given the  uncertainties
in  different simulations at ultra high energies,  
%In such case the point at 57 \tev in this figure would be 
 the estimate for ${\cal R}_{\rm el}$  at 57~\tev\ could then be lowered (closer to the TOTEM value), thereby  opening the possibility that   an  asymptotic plateau  for \sigeltosigtot \   may  have already been reached. In fact ${\cal R}_{\rm el}$ has been rising from lower energies until  the value $\sim 1/4$ at the TeVatron, and at  LHC7. Data at 8 \tev\ could indicate whether   we have reached a plateau or not.

 \section{The two exponential model at LHC energies}
In order to perform a test of the ASR in a model independent way, we have analyzed the data for the differential elastic cross section through a  parametrization of the scattering amplitude, proposed by Barger and Phillips (BP) in 1973~\cite{bp1973}. 
%In the next section we shall see that this simple parametrization describes quite well the LHC data and that the data indicate satisfaction of the sum rules. We shall then apply  the parametrization to lower energy data, from ISR onwards, and study the energy behaviour of the amplitude and the approach to asymptotia, from the sum rules. We   make predictions for expected  data measurements  at higher LHC energies.

%\section{Model independent description of the elastic scattering from ISR to LHC}
In 1973, 
%Barger and Phillips 
BP 
proposed 
%In what they called a model independent analysis of \pp \ scattering~\cite{bp1973}.
%Phillips:1974vt}. 
%They proposed 
%Barger and Phillips proposed 
 two different parametrizations, consisting in using a  phase
%, which turns out to be energy dependent, 
and two exponentials to describe  the elastic scattering amplitude as a function of  momentum transfer in the range $-t=0.15\div5.0\ \gev^{2} $.
% and both inspired by Regge behaviour for the shrinking of the diffraction peak. The two  expressions are similar, in the sense of describing the $t$-dependence of the amplitude as the sum of two exponentials and a relative phase. 
The first parametrization was  
%a model independent parametrization,  
given as
\bea
{\cal A}(s,t)= i\left[
\sqrt{A(s)}e^{\frac{1}{2}B(s)t }+ \sqrt{C(s)}e^{i\phi(s)} e^{\frac{1}{2}D(s)t}\right] \,,
\label{eq:bpeq1-ampl}\\
\frac{d\sigma}{dt}=A(s)e^{B(s)t}+C(s)e^{D(s)t}+2 \sqrt{A(s)}\sqrt{C(s)}  e^{(B(s)+D(s))t/2} 
\cos{\phi}\,,\label{eq:bpeq1-1}
\eea
with five $s$-dependent real parameters, $A,\ B,\ C, \ D,\ \phi$.
The total cross section in this model is given as
 \bea
 \sigma_{\rm total}=4\,\sqrt{\pi}\,\Im m {\cal A}(s,t=0)=4\,\sqrt{\pi}\,
\left[\sqrt{A(s)}+\sqrt{C(s)}\cos\phi\right]\,.
 \label{eq:sigtot}
 \eea
Using the above expression,
%, but with $A(s)$ and $B(s)$ kept fixed throughout the energy range they considered in this model, 
BP fitted data from $p_{\rm lab}=12\ \gev$ to $1496~\gev$ ($\sqrt{s}\approx 5\div53\ \gev$).  The values of the parameters at $\sqrt{s}=53\ \gev$ are given as $\sqrt{A}=6.55,  \sqrt{C}=0.034 $ in 
$\sqrt{\mb/\gev^2}$ and $ B=10.20, D=1.7$ in $ \gev^{-2}$, $ \phi=2.53 \ rad$.  
The first exponential was  seen to have normal Regge shrinking, namely
\bea
\label{B}
B=B_0+2\alpha' \ln s\,, \ \ \ \ \ \ \ \ \ \alpha'\approx 0.3\,,
\label{eq:bpxB}
\eea
while the second exponential term appeared to be  constant in this energy range. 
With this parametrization,  the phase $\phi$ was  fitted to be always larger than $\pi/2$, so that the interference term was  always negative. The second parametrization interpreted the
 energy dependence of the parameters in terms of Regge-Pomeron exchange and  fits 
to data were as good as in the first case, but  more model dependent. However, 
%the energy dependence given by 
Eq.~(\ref{eq:bpxB}) for  the forward slope $B$ poses now a problem, as recently discussed in {Ref.}~\cite{ryskin}, since  TOTEM  data can  be described by Eq.~(\ref{eq:bpxB}) only if  $\alpha' $ is not a constant, but has  a logarithmic energy dependence. On the other hand, by relinquishing any model interpretation, one can use the model of  Eq.~(\ref{eq:bpeq1-ampl}) as it appears to be  the  simplest possible way to describe the $t$-dependence of  \pp\ elastic scattering amplitude from ISR onwards.

The simplicity of the  model in Eq.~(\ref{eq:bpeq1-ampl}) suggests to use it  to describe present LHC data.  Reading the TOTEM data from {Ref.}~\cite{totem-elastic}, we show the  result in the left panel of Fig.~\ref{fig:dsigellhcfits}. We find that two exponentials and a phase can  give a very good description of TOTEM data,  with  the tail    well described 
%in agreement by an exponential with the two exponential model, 
by  a term  $e^{D t}$, with $B/D\sim 4$  and a  relative weight of the two terms at LHC7  $\sqrt{A/C}\sim22 $. With this parametrization, the position of the dip (which the model fit gives  at  $-t=0.52 \ \gev^2$) and the behaviour around the dip are both   well reproduced.  This result differs  from  the proposal in {Ref.}~\cite{DL2012} and also from the suggestion by TOTEM~\cite{totem-elastic} that  the tail,  past the dip at $-t=(0.53 \pm 0.01^{\rm stat}\pm 0.01^{\rm syst})\, \gev^2$, is described as being compatible with a behaviour  of the type  $|t|^{-n}$ with $n=7.8 \pm 0.3^{\rm stat}\pm 0.1^{\rm syst}$.  The TOTEM suggested behaviour  is close to a description of the tail through  a   QCD type  contribution from  three gluon exchange advanced in {Ref.}~\cite{DL1996}. It is also close to the observation
 %e  attempted a different description, mindful of the fact that .
%At the same time, 
in {Ref.}~\cite{bp1973}, 
%it is also suggested that
that  the ``second exponential'' can be identified with a term proportional to $p_T^{-14}$, valid, according to the authors, for all available $pp$ data for $s>15\ \gev^2$. Following these suggestions, and mindful of the fact that the form factor dependence of the amplitude would contribute 
%like 
to the cross section with a term as $(-t)^{-8}$, we have tried a slight modification of the 
%Barger and Phillips 
BP model, which  consists in substituting the second term in Eq.~(\ref{eq:bpeq1-ampl}) with a term proportional to the {  2$^{\rm nd}$ power} of the proton form factor,  namely to write
 \bea
 {\cal A}(s,t)=i\left[\sqrt{A(s)}e^{Bt/2}+\frac{\sqrt{C(s)}}
{(-t+t_0)^4} e^{i\phi}\right]\,.
\label{eq:ffmodel}
%\frac{d\sigma}{dt}=|\sqrt{A(s)}e^{Bt/2}+\frac{\sqrt{C(s)}}
%{(-t+t_0)^4} e^{i\phi}|^2 
\eea}
%{\it i.e.}
%\bea
%\frac{d\sigma}{dt}=Ae^{Bt}+\frac{C}{(-t+t_0)^8} +2\cos{\phi}\frac{\sqrt{A}\sqrt{C}e^{Bt/2}}{(-t+t_0)^4}\,.
%\label{eq:bpours}
%\eea
{ With   $t_0$  a free parameter}, this allows a comparison 
with the parametrization of the tail in {Ref.}~\cite{totem-elastic}, 
but the fit worsens, and the tail 
%where the second term is inspired by the possible appearance of a form factor or of perturbative QCD  effects. Using a model such as
%\be 
%{\cal A}(s,t)=i[
%\sqrt{A(s)}e^{B(s)t/2}+
%\frac{For small values of $$
%\sqrt{C(s)}e^{i\phi(s)}
%}{
%(-t+t_0)^4}]
%\ee
%with $t_0$ as a free parameter, we find that the fit worsens, since the tail i
is everywhere better described by an exponential. We show this fit in the right hand panel 
of Fig.~\ref{fig:dsigellhcfits}.
\begin{figure}[h!]  
\centering
\hspace{-1cm}\resizebox{0.5\textwidth}{!}{
\includegraphics{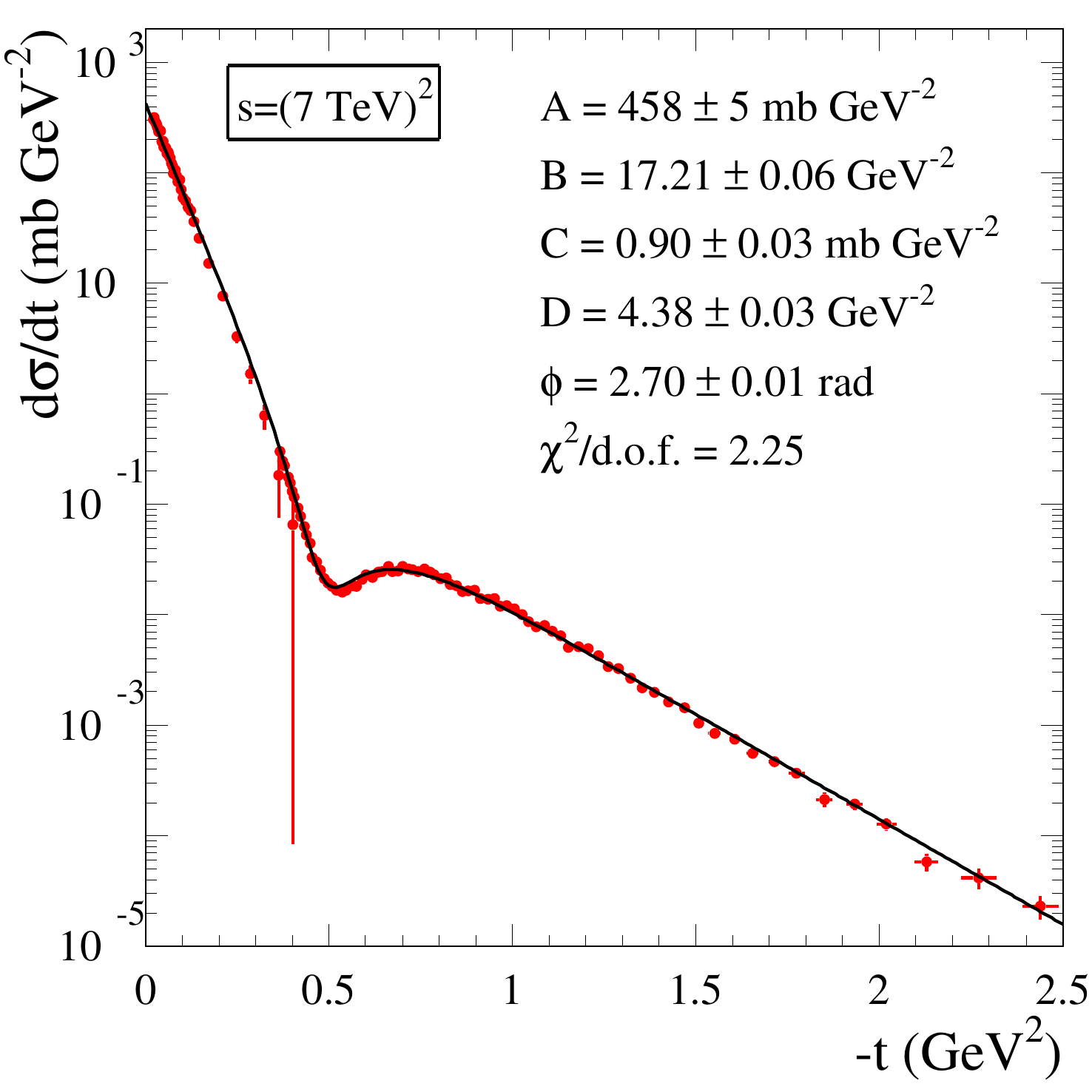}}
%dsigdt_el_lhc_expexpfit_5param}}
\resizebox{0.5\textwidth}{!}{
\includegraphics{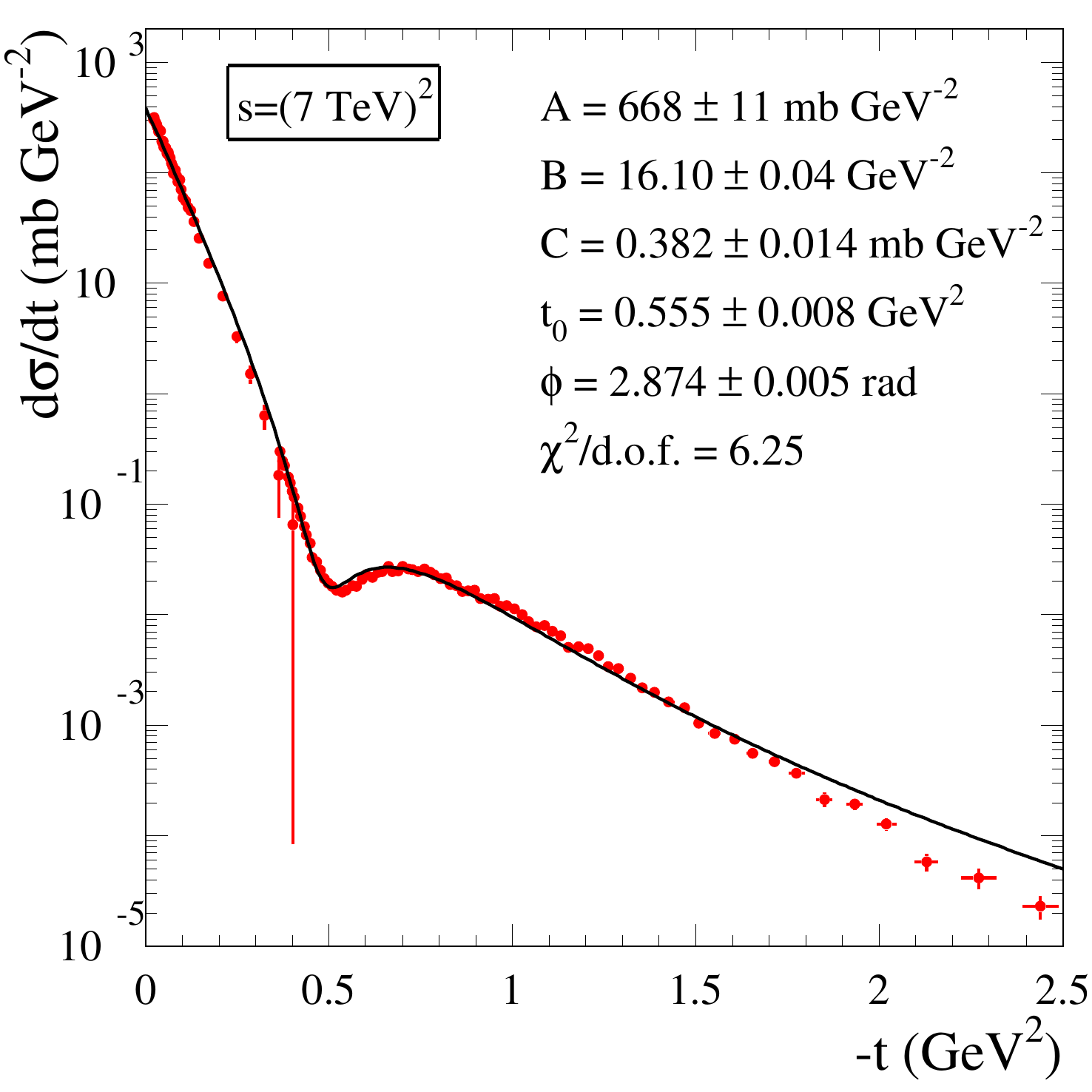}}
%totem-exp-ff}}
%dsigellhcFFfit}
\caption{\label{fig:dsigellhcfits} 
%Kumac 5
 Five-parameter fit of   $pp $ data for the elastic differential cross section at LHC7. At left the fit  uses  the two exponential model from {Ref.}~\cite{bp1973}, at right   a form factor term is used to describe the tail, according to Eq.~(\ref{eq:ffmodel}). Parameter values in the two cases are indicated. TOTEM data were read from {Ref.}~\cite{totem-elastic}. }
\end{figure}
From here on, we shall use the two exponential model and proceed to check the ASR.
Since the phase $\phi\ne \pi$, for  the satisfaction of both ASR the model of Eq.~(\ref{eq:bpeq1-ampl}) needs to be modified so  as to include a real part in the leading amplitude $i \sqrt{A(s)}$,  { {\it i.e.}  the leading term (at $t\approx 0$)} 
%in Eq.~(\ref{eq:bpeq1-1})  } 
should acquire an extra phase.
%, i.e. \be i \sqrt{A(s)}\rightarrow \sqrt{\pi} A_Y =
%\sqrt{\pi} [a_R(s)+ia_I(s)]=
%i \sqrt{\pi} a_I(s)[1+\frac{a_R(s)}{a_I(s)}]  i \sqrt{\pi} a_Y(s)[1+\frac{\Re e A_Y(s,0)}{\Im m A_Y(s,0)}]\sim i
 % \frac{\sigma_{total}}{4\sqrt{\pi}}[1+\rho(s,0)]
%\ee
%where the last equality neglects the (small) contribution of the second term in Eq.~(\ref{eq:bpeq1-1}) at %$t=0$.
%Thus,
%A  real part can be added to the first term  if one has a model for $\rho(s,0)$ . Such 
%A value for $\rho(s,0)$ 
Rewriting  the amplitude of Eq.~(\ref{eq:bpeq1-ampl}) as
\bea
{\cal A}(s,t)=\left[ i{\hat A_I}+{\hat A_R}\right]e^{\frac{1}{2}B(s)t }+\left[ 
i\sqrt{C(s)}\cos \phi -\sqrt{C}\sin \phi \right] e^{\frac{1}{2}D(s)t}\,,
\label{eq:bpampl}
\eea
%{where $\hat A_I$ and $\hat A_R$ stand for  imaginary and real parts of $i\sqrt{A}$},
one has 
\bea
\rho(s,0)=
\frac{
{\hat A_R}-\sqrt{C}\sin\phi }
{{\hat A_I}+\sqrt{C}\cos\phi }\,.
\eea
%With this, at LHC we  obtain $\rho(s,0)=0.134 \div 0.115$.  Given  the uncertainty in this determination, and given the fact that the second term in Eq.~(\ref{eq:bpours}) is smaller than the first one  by a factor of 20,  
We use the  result for the leading contribution to $\rho(s,t=0)$ to give a real 
part to the first term, {\it i.e.} 
\bea
\hat \rho(s,t=0)=\frac{\hat A_R}{\hat A_I}= \frac{\pi}{2p \ln(s/s_0)}.
\eea
With this,  we now find 
\bea
SR_0=\sqrt{\frac{A}{1+{\hat \rho}^2}}
\,\frac{1}{\sqrt{\pi}B}{\hat \rho} -
\,\frac{\sqrt{C} \sin{\phi}}
{\sqrt{\pi}D}
=0.045\div0.066\,,
\label{eq:s0numbers}
\eea
for $p=0.77 $ and $0.66$ respectively, and 
\bea
SR_1=\sqrt{\frac {A}{1+{\hat \rho}^2}}
\,\frac{1}{\sqrt{\pi}B}+
\,\frac{\sqrt{C}}
{\sqrt{\pi}D}\cos{\phi}=0.94\,,
\label{eq:s1numbers}
\eea
with  $A={\hat A}_R^2+{\hat A}_I^2$,  and inserting the parameter values from the figure, with  $\pi/2<\phi<\pi$. One observes  a good  satisfaction of both ASR. 
Notice however, that, in general,   the conjecture  for $\rho(s,0)$ proposed in  Eq.~(\ref{eq:rho}) cannot be valid at lower energies, because  the approximation for the amplitude in Eq.~(\ref{eq:phase block}) 
% and the approximation for the total cross section
 is only valid at asymptotic energies. Thus, for  uniformity with the ISR case, curves in Fig.~\ref{fig:dsigellhcfits} have been done with ${\hat A}_R=0$.  
 %At such energies,  the parameter $\rho$ in this model is given as
Applying the above model to the ISR data for $pp$ scattering at $\sqrt{s}=53\ \gev$ \cite{isr}, we obtain the fit shown in the left panel of Fig.~\ref{fig:isr5param}.
\begin{figure}[h!]
\centering
%\vspace{-1cm}
\hspace{-0.8cm}
\resizebox{0.48\textwidth}{!}{
\includegraphics{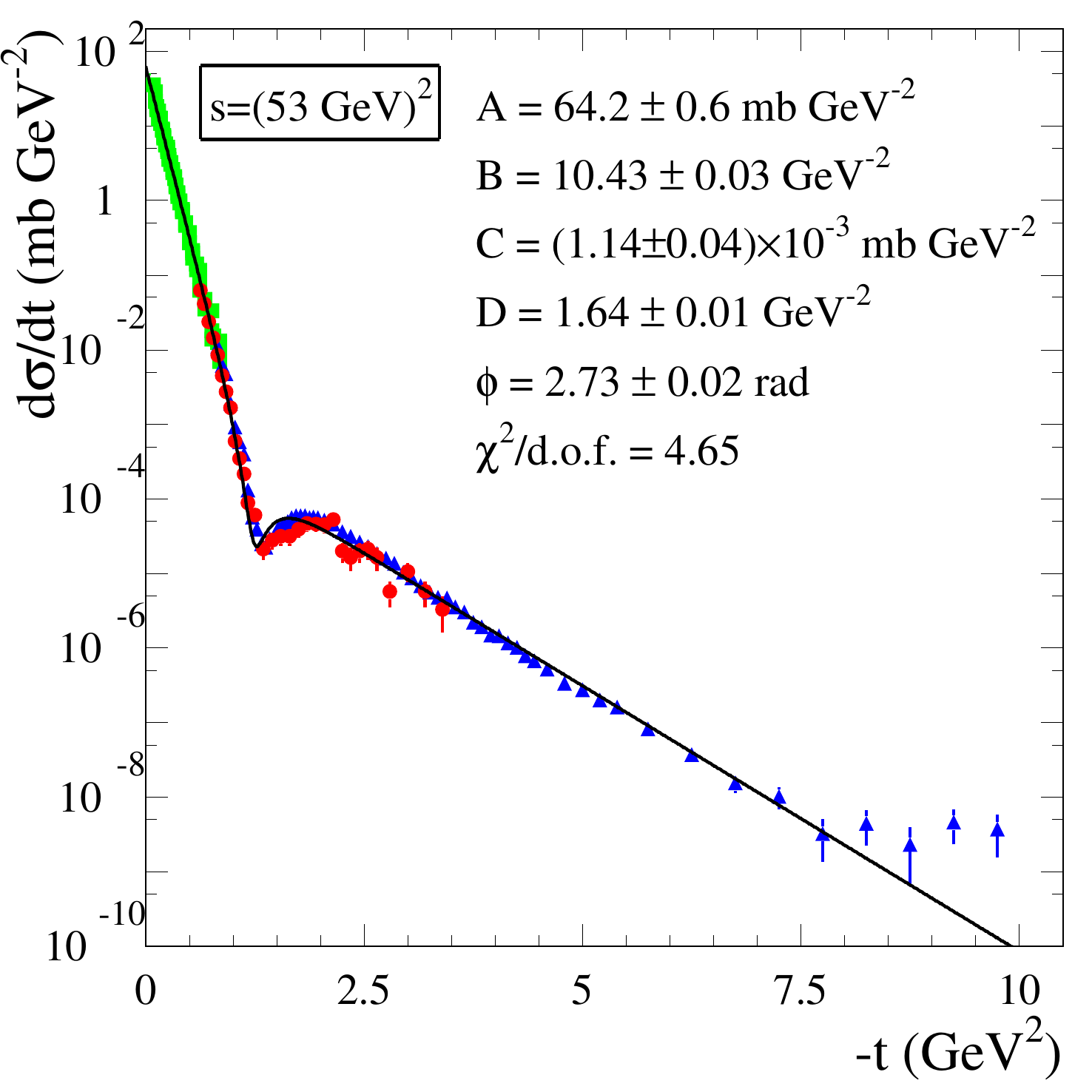}}
\hspace{0.5cm}
%dsigdt_el_isr_expexpfit_5param_bis}}
\resizebox{0.5\textwidth}{!}{
\includegraphics{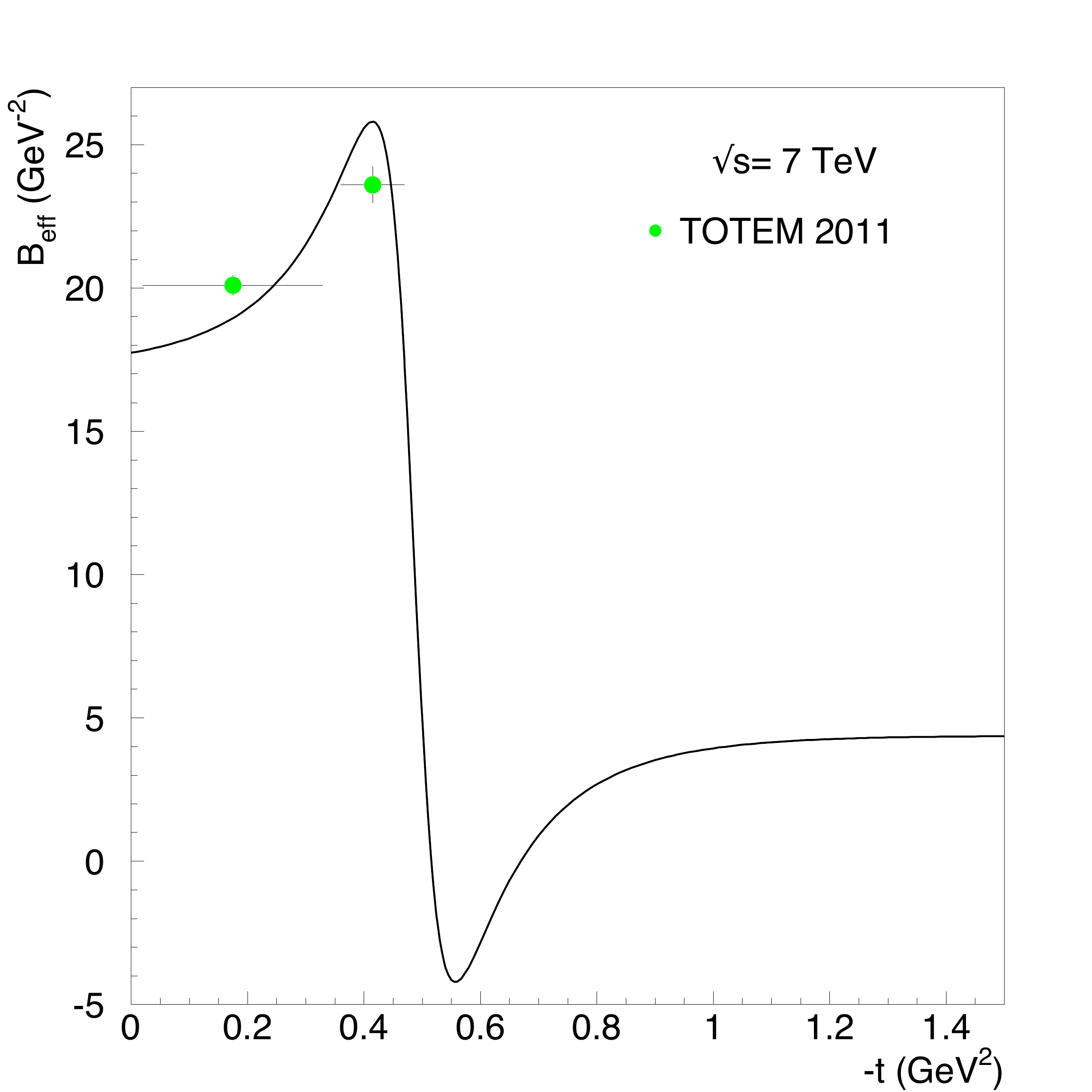}}
\caption{\label{fig:isr5param} At left, the description of $pp$ scattering data at ISR, using the model of Eq.~(\ref{eq:bpeq1-1}). A five-parameter fit gave the indicated values for the parameters. At right we show the $t$-dependence of the effective slope at LHC7 from Eq.~(\ref{eq:beff-expression}).
%. {\it ricordarsi di cambiare o controllare i valori di s0 e s1 per ISR}
}
\end{figure}
 We notice that the application of the ASR to $pp$ scattering at these lower energies,  gives 
$SR_1\approx 0.7$ for  the parameter values obtained by  the fit. As expected,  the ASR for 
the imaginary part of the amplitude in $b$-space is  not yet satisfied at ISR.   
As for the second ASR at ISR,
 % if we used  Eq.~(\ref{eq:rho}), $s_0=0.2$}. {\bf As mentioned 
 the expression for the real part of the amplitude as, discussed in Sec.~2, 
cannot be applied here, because at ISR we are far from an asymptotic regime.
  %. \sout{Thus we can only say that if we estimate the sum rule using only  the  real part of the amplitude coming   from the second term in Eq.~(\ref{eq:bpeq1-ampl}), one would obtain $s_0=-7\ 10^{-3}$. } }
Before leaving this discussion, we point out that   the value obtained for the
    total cross section in this model  is  below the experimental value. This is particularly true for  ISR energies, where
    %, inserting the  parameter values from 
    the fit  gives $\sigma_{\rm total}=(35.6 \pm 1.7)\,\mb$   to be compared with  values with very small errors,  such as $(42.71\pm 0.35)\,\mb$~\cite{amaldi}. At LHC7 the total cross section  is better reproduced by the fit, with a value of $(91 \pm 5)\,\mb $, although still a bit low. From Figs.~\ref{fig:dsigellhcfits} and \ref{fig:isr5param}, we conclude that this model is approximately valid at lower energies, better apt to describe data at LHC7 and onwards.
 
%cannot be valid, we cannot say much, except that  $s_0=-0.005$ with ${\hat A}_R=0$.}
 \section{The slope parameter  in the two exponential model}
 TOTEM released values for $\sigma_{\rm total}$, $\sigma_{\rm elastic}$, the slope $B_{\rm exp}$ and the position of the dip.   
 %at $-t=(0.53\pm 0.01^{\rm stat}\pm 0.01^{\rm syst}) \,\gev^2$. %At LHC we have seen that 
 The total cross section  is reproduced by the five-parameter fit ($\sigtot$ is let unconstrained) of  the two exponential model within 10\%, and we  have a good determination of the position of the dip,  which,  in our fit, occurs at $-t_{\rm dip}=0.52\ \gev^2$.  
 
As for the slope parameter, one cannot immediately  compare the value of this model parameter $B$ with TOTEM data, where it is   defined    as
\bea
\frac{d\sigma_{\rm el}}{dt}=
\left.\frac{d\sigma_{\rm el}}{dt}\right|_{t=0}\ e^{B_{\rm exp}t}
\label{eq:bexp}
\eea
and is measured in two $t$-intervals, with  $B_{\rm exp}=(20.1 \pm 0.2^{\rm stat}
\pm 0.3^{\rm syst})\, \gev^{-2} $ for  smaller $|t|$ values, 
{\it i.e.} $0.02 < -t< 0.33\, \gev^2 $, and $B_{\rm exp}=(23.6\pm 0.5^{\rm stat}\pm 0.4^{\rm syst})\, \gev^{-2}$ from  the interval $0.36 < -t< 0.47  \ \gev^{2}$. Such difference is well understood in the two exponential model.  With  the definition for the effective slope \cite{blockcahnrep}
\bea
 B_{\rm eff}(s,t)\equiv
\frac{d}{dt}\ln\left(\frac{d\sigma_{\rm el}}{dt}\right)\,,
 \label{eq:beff}
 \eea
one obtains  a function changing with $t$, {\it i.e.}
% the effective slope % in the region of  the TOTEM measurement is 
%is not a constant, and its value, at any given $t$, 
%can be obtained as
\bea
B_{\rm eff}(s,t)=\frac{ABe^{Bt}+CDe^{Dt}+ \sqrt{A}\sqrt{C}(B+D)e^{(B+D)t/2} \cos\phi }{\dsigdt} \,.
\label{eq:beff-expression}
\eea
Using the parameter values from the left hand panel of Fig.~\ref{fig:dsigellhcfits},  we show in the right-hand panel of Fig.~\ref{fig:isr5param} the $t$-dependence of the slope $B_{\rm eff}$ at LHC7, compared with the values  { given} by the TOTEM experiment in the two different $t$ intervals, whose extension is indicated by the horizontal bars.
At $t=0$, one finds 
%of Table~\ref{tab:kumacfits5param},
%, \ref{tab:kumacfits4param} and  \ref{tab:eyefit},  we find 
%\bea
\bea
 B_{\rm eff}(7\ \tev,0)=\left.\frac{d}{dt}\ln\left(\frac{d\sigma_{\rm el}}{dt}\right)\right|_{t=0}=17.7 \ \gev^{-2}\,, \ \ \ \ \ \ \mbox{(five-parameter fit)}\,.
\eea
 Notice that the figure is obtained by neglecting the contribution from ${\hat A}_R$. Clearly this approximation is not valid where the real or the imaginary part of the amplitude  approaches zero.  It is possible to identify the two $t$-values where this will happen
%the slope will change 
before the dip, namely where the real part of the amplitude is zero, and a subsequent one, close or very close to the dip,  where the imaginary part is zero. At LHC7, these two points occur at
\bea
-t_R&\!\!\!=\!\!\!&\frac{2}{B-D}\ln\left(\frac{\hat{A}_R}
{\sin \phi \sqrt{C}}\right)=
0.28\div 0.30\ \gev^2\,,\\
-t_I&\!\!\!=\!\!\!&\frac{2}{B-D} \ln
\left(
\frac{\sqrt{A/C)}}{|\cos \phi|}
\right) =0.5 \ \gev^2\,,\label{eq:tim}
\eea
where the uncertainty in $t_R$ depends on the value for ${\hat \rho}$ and we have 
neglected ${\hat \rho}^2$ in Eq.~(\ref{eq:tim}) and, again,  we have used the 
parameter values indicated in  the left hand panel of Fig.~\ref{fig:dsigellhcfits}.
%Fig.~\ref{fig:isr5param}.

An asymptotic value for the slope  $B_{\rm eff}(s,t)$, can be obtained 
through the ASR, already  well satisfied at LHC7. Thus, 
in the exact limits $SR_0=0$, $SR_1=1$, for  asymptotic energies
% such as 
at LHC and beyond,
one obtains for the effective slopes
 \bea
 B_{\rm eff}(s, t_R)&\!\!\!=\!\!\!&\frac{\sigma_{\rm total}}{4 \pi}\,,\\
 %=20.2\  \gev^{-2}\ \ \ \ \ -t_R=0.21 \ \gev^2\ \ \ \sqrt{s}=7\ \tev\\
 B_{\rm eff}(s,t_I)&\!\!\!=\!\!\!&B+D- \frac{\sigma_{\rm total}}{4 \pi}\,.
 %=1.6\ \gev^{-2} \ \ \ -t_I=0.502 \ \gev^2\ \ \ \sqrt{s}=7\ \tev
 \eea
 We find, in agreement with TOTEM, that { past $t_R$,} where the real part of the amplitude is zero, the slope is $\sim 20\  \gev^{-2}$, and that it increases to higher values, to then start decreasing} as it approaches  $t_I$ where the imaginary part goes to zero {  in correspondence with the occurrence of the dip.

We  note  that the ASR imply that the slope $ B_{\rm eff}(s,t_R)$ from the two exponential model should grow 
%like 
as $\sigma_{\rm total}$, thus
 %like 
 as $\ln^2 s$, if the  Froissart bound is saturated, or as $\ln^{1/p}s$ for a slower rise.
We see that the same behaviour is also true asymptotically for the (constant in $t$) leading  parameter $B$ of the two exponential model, for which we also have
 the well known result
 \bea
 {\cal R}_{\rm el}(s) =
\frac{\sigma_{\rm elastic}(s)}{\sigma_{\rm total}(s)}\simeq \frac{\sqrt{A}}{4\sqrt{ \pi} B}
\approx \frac{\sigma_{\rm total}}{16 \pi B}
\eea
 recently also discussed in {Ref.}~\cite{fagundes} on general grounds.}

We conclude that, if the two exponential model is a good representation of the amplitude { at LHC} in the range $0\le -t \le 3\ \gev^2$, and if the total cross section rises faster than $\ln s$,  the slope of the leading term in this model 
%defined from Eq.~(\ref{eq:bexp})
should grow faster than $\ln(s/s_0)$.
% sout{, a result in conflict with the Regge pole theory as given in 
%Eq.(\ref{B}).}
Notice that in {Ref.}~\cite{ryskin},  an analysis of the effective slope from NA8 to TOTEM seems to indicate a rise according to $\ln^2 s$. Thus, just as  in the case of the total cross section, the question of the energy dependence of the effective slope is still very much open. %Precision data at $8\ \tev$ would 

In conclusion, we stress the following points concerning the slope parameter:
%\subsection {Some general remarks about the '' slope'' parameter by Yogi }
\begin{enumerate}
\item %As observed above,
 One can not speak about a unique slope if there is more than one term with different $t$ dependence in the elastic amplitude, as required by the presence of a dip, 
%\par\noindent
%2. And the TOTEM data at 7 \tev with a dip
 in the differential cross section.
 \item 
 Even in the region before the dip, the necessity of  {\it at least} two terms, whose interference gives rise to the dip, is felt and  TOTEM itself  provides  two different values for the slope in two different regions of $t$.
Hence
%, until an energy high enough -if ever- is reached where there are no such dips, so that there remains one leading term, 
it is not meaningful to 
%talk about 
{consider only} a single slope and its asymptotic behavior. In a two component model such as the one considered above, all one can discuss is the relative growth of the two slopes with energy. 
\item
 Even in the rather simple model analyzed here, we find that the slope 
%neat 
near
$t=0$ is less than that around $t= - 0.2 \ \gev^2$, which is less than that around $t= - 0.4 \ \gev^2$ and hence the question of the growth with energy becomes a strong function of the $t$ range under consideration.
\end{enumerate}
{ The curve in the right hand panel of Fig.~\ref{fig:isr5param} shows that at LHC this model predicts an increase of the slope parameter in the range $0 <-t\lesssim 0.4 \ GeV^2$. This is in contrast with the behaviour observed at lower energies, notably at ISR for very small $t-$values \cite{Ambrosio1982,Amos1985}. Notice however that our findings at LHC are in agreement with those expected in Ref.~\cite{Desgrolard1997} at asymptotic  energies.}

\section*{Conclusion}
We have 
%presented 
utilized two asymptotic sum rules  derived earlier for the elastic scattering amplitude in $pp$ scattering \cite{Pancheri:2004xc} and have tested them using  a well known parametrization of the elastic differential cross section with two exponentials with a relative phase. Our study of   TOTEM data at LHC7 finds  that this model can describe well the data before and past the observed dip. Similar conclusions about the behaviour past  the dip, have also been advanced in {Ref.}~\cite{troshin}.

{ A possible interpretation of this model  is that the exponential behaviour exhibited both before and after the dip corresponds to a resummation of soft terms accompanying the  leading  $C=+1$, two gluon exchange, and $C= -1$,  three gluon exchange term, with non leading terms giving rise  to  $\phi\neq \pi$.}

Given { the   general nature of } the parametrization discussed here, we recommend its usage for further studies of  the behaviour of the scattering amplitude at higher energies.  
\section*{Acknowledgement}
We thank E. Lomon and R. Godbole   for useful suggestions and enlightening discussionss about { elastic scattering } data. One of us is grateful to the Center for Theoretical Physics of MIT for hospitality.
% while this work was prepared.   
Work partially supported by 
Spanish MEC (FPA2006-05294, FPA2010-16696) and by
Junta de Andalucia (FQM 101). This work has been supported in part by the  Spanish Consolider Ingenio 2010 Programme CPAN
(CSD2007-00042).YS would like to thank his colleagues on the Auger Collaboration for discussions.
%{\it  Others?} 
 
%We use then the values and trends exhibited by the parameters of the {\it eye-fit} to make two predictions for the elastic differential cross section at LHC higher energies, which shown in Fig.~\ref{fig:dsigdtBPppLHC8-14}.   
% Agnse says: far vedere che e' possibile fittare i dati in modo ragionevole (almeno ad occhio con questa funzione a due esponenziali e avere a lo stesso tempo un comportamento con l'energia dei diversi parametri che ha un senso> (perche quelli del kumac non ce l'ha)
%\paragraph{Comparison with a form factor model}

%  \section{Asymptotic Sum Rules and the two exponential model}
%\section{Scattering amplitude in b-space and comparison with black disk models}
%\section{Adding an asymptotic real part}

\end{document}